# Experimental demonstration of separating the wave–particle duality of a single photon with the quantum Cheshire cat

Jia-Kun Li,[1,2,3] Kai Sun,[1,2,3,*] Yan Wang,[1,2,3] Ze-Yan Hao,[1,2,3] Zheng-Hao Liu,[1,2,3] Jie Zhou,[4] Xing-Yan Fan,[4] Jing-Ling Chen,[4,†] Jin-Shi Xu,[1,2,3,‡] Chuan-Feng Li,[1,2,3,§] and Guang-Can Guo[1,2,3]

[1]CAS Key Laboratory of Quantum Information, University of Science and Technology of China, Hefei 230026, China

[2]CAS Center for Excellence in Quantum Information and Quantum Physics, University of Science and Technology of China, Hefei 230026, China

[3]Hefei National Laboratory, University of Science and Technology of China, Hefei 230088, China

[4]Theoretical Physics Division, Chern Institute of Mathematics, Nankai University, Tianjin 300071, China

**Abstract**: As a fundamental characteristic of physical entities, wave–particle duality describes whether a microscopic entity exhibits wave or particle attributes depending on the specific experimental setup. This assumption is premised on the notion that physical properties are inseparable from the objective carrier. However, after the concept of the quantum Cheshire cats was proposed, which makes the separation of physical attributes from the entity possible, the premise no longer holds. Furthermore, an experimental demonstration of the separation of the wave and particle attributes inspired by this scenario remains scarce. In this work, we experimentally separated the wave and particle attributes of a single photon by exploiting the quantum Cheshire cat concept for the first time. By applying a weak disturbance to the evolution of the system, we achieve an effect similar to the quantum Cheshire cat and demonstrated the separation of the wave and particle attributes via the extraction of weak values. Our work provides a new perspective for the in-depth understanding of wave–particle duality and promotes the application of weak measurements in fundamentals of quantum mechanics.

---

[*] ksun678@ustc.edu.cn
[†] chenjl@nankai.edu.cn
[‡] jsxu@ustc.edu.cn
[§] cfli@ustc.edu.cn

**Introduction**

Waves and particles are considered two fundamental attributes of light and matter. Based on optical phenomena, such as interference, diffraction, and scattering, light exhibits wavelike behaviour, whereas according to other phenomena, such as light travelling in straight lines and the photoelectric effect, light behaves like a particle[1]. The debate on whether light is a wave or a particle has lasted for hundreds of years since the 17th century[2-4].

Wave–particle duality is probably one of the most intriguing counterfactual concepts in quantum mechanics, in which the interpretation of the wave and particle attributes of an objective entity is quite different from those in classical world. In Niels Bohr's complementary principle[5], detecting the wave–particle duality of light depends on the devices. As illustrated in a traditional Mach–Zehnder interferometer (MZI), the presence or absence of the second beam splitter (BS) will determine whether the wave or particle attribute of light is observed[6], while two properties cannot be simultaneously observed. However, after Wheeler's "delayed-choice experiment"[7, 8] and its recent quantum versions[9], quantum wave–particle superposition, in which both the wave and particle attributes are observed simultaneously, has been realized in experiments, including those using a single photon[10, 11] and two entangled photons[12]. Recently, several outstanding works exploring profound connotation in wave–particle duality have been conducted, including the investigation of the linear form of duality relation with asymmetric beam interference[13], the experimental progress in the large-scale quantum nanophotonic chip[14], and theoretical achievements with the electron[15]. Although a series of outstanding theoretical and experimental works that deepen the understanding of wave–particle duality have been reported, it remains unclear which path a single photon (or any other particle) takes when it enters the MZI (double-slit setup). Moreover, there is still considerable long-term debate regarding this puzzle[16-18].

In recent years, the concept of weak measurement[19-24] has provided new ideas for addressing this challenge. Due to its characteristic of reducing the disturbance caused by measurement, some information of a quantum state can be extracted without collapsing the state into the eigenstate[25]. Numerous experiments based on weak measurements have been carried out for various applications, such as the observation of the spin Hall effect of light[26], direct measurement of quantum wavefunctions[27], and testing of the violation of Bell's inequality[28]. In particular, under the weak measurement framework, the quantum Cheshire cat effect[29], inspired by the famous novel, "Alice in Wonderland[30]," can be achieved through appropriate pre- and post-selections. The quantum Cheshire cat concept reveals an unconventional phenomenon: physical properties can be separated from the original object. In quantum mechanics, this indicates that the intrinsic property of a particle, such as its spin, can be disembodied from the particle itself. Such a separation has been demonstrated experimentally in neutron[31] and optical systems[32].

During the process of proposing the quantum Cheshire cat theory, initially regarding the strange "separation" phenomenon as just an optical illusion[29] due to the interference of measurements with each other under the framework of measurement theory in quantum mechanics is tempting. Consequently, the results of separate measurements are inconsistent with those of simultaneous measurements. However, weak measurement theory proves that a paradox does exist; that is, physical properties can be separated from their carriers, which can help explain several paradoxes in quantum mechanics[33-35].

Generally, in quantum mechanics, for both strong and standard weak measurements, an additional auxiliary pointer is required to read out the measurement results[36]. However, as the system scales, the number of pointers increases, significantly magnifying the overall complexity of the system. This explains the lack of universal operability, attributed to the considerable difficulty associated with experiment. With progress in both theory and experiments, a relatively simple scheme for extracting weak values, called imaginary-time evolution (ITE), has been proposed[25, 37, 38]. In this work, we adopt this simple method to extract the weak values.

Based on the quantum Cheshire cat concept and the introduced weak measurement technique, it is possible to separate the wave and particle attributes of a single particle, which provides new insights for addressing several fundamental challenges, such as Young's double-slit experiment with either the particle or wave attribute. Furthermore, it is interesting to investigate the photoelectric effect with only the wave attribute of photons and whether interference, diffraction, and other phenomena reflecting the fluctuation of light can be observed with only the particle attribute. To answer these fundamental and intriguing questions, the prerequisite step is to successfully separate the wave and particle attributes of a single particle.

Recently, a thought experiment was proposed to separate the wave and particle attributes of a quantum entity[39]. Following the theoretical conception, here, we experimentally demonstrate the separation of the wave and particle attributes of a single photon based on the quantum Cheshire cat concept for the first time. By choosing proper pre- and post-selected states and implementing the weak measurement strategy, we successfully extract weak values of different operators, and the results provide solid support confirming that the wave and particle attributes have indeed been spatially separated. Our work motivates further thought to wave–particle duality and facilitates the investigation of the fundamentals of quantum mechanics with the weak measurement method.

**Results**

*Theoretical framework.* The schematic of the wave–particle duality separation is shown in Fig. 1. First, through the wave–particle toolbox[12, 39], the superposition state of the wave and particle attributes ($|\psi\rangle = \cos\alpha|Particle\rangle + \sin\alpha|Wave\rangle$) is prepared. Next, as mentioned above, a quantum Cheshire cat represents a counterfactual phenomenon: the separation of the physical property from the carrier. To illustrate this phenomenon, we employ an improved MZI with the appropriate pre-selection and post-selection setups. It is supposed that a cat passes through the MZI, and a surprising fact emerges. The cat (wave attribute) goes through one path of the interferometer, whereas its grin (particle attribute) goes the other way, as shown in Fig. 1**a.** By observing the weak values defined by the pre- and post-selection states and the specific observable, $\hat{A}$, we can extract the weak value of this observable:

$$\langle\hat{A}\rangle_w = \frac{\langle\psi_f|\hat{A}|\psi_i\rangle}{\langle\psi_f|\psi_i\rangle} \tag{1}$$

Here, the pre-selection state is set as $|\psi_i\rangle = (|L\rangle + |R\rangle)(\cos\alpha|Particle\rangle + \sin\alpha|Wave\rangle)/\sqrt{2}$, which can be obtained after the quantum state $|\psi\rangle$ passes BS1 (see Fig. 1b); $|L\rangle$ and $|R\rangle$ corresponds to the two paths of the interferometer.

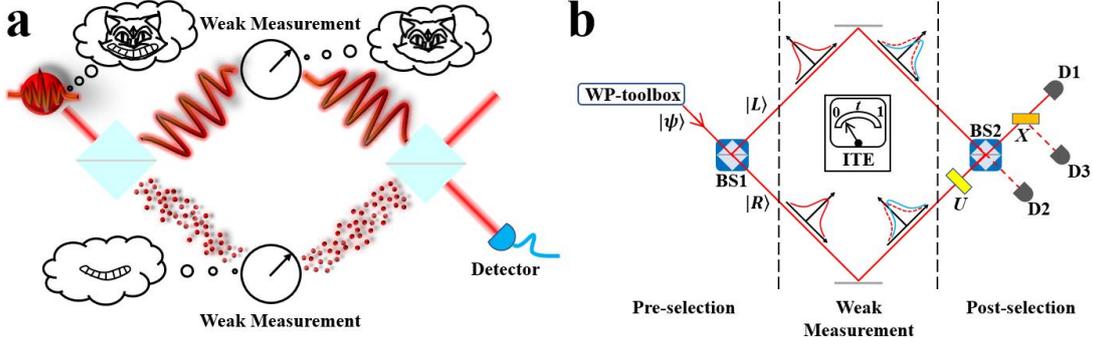

**Fig. 1. Schematic for the separation of the wave and particle attributes, inspired by the quantum Cheshire cat concept. a** Conceptual diagram. A cat (wave–particle superposition state) enters the interferometer. By extracting weak values, it is surprising to find that the cat (wave attribute) and its grin (particle attribute) have been separated spatially. **b** Theoretical scheme. The wave–particle superposition state, $|\psi\rangle$, obtained through the WP-toolbox passes the beam splitter (BS1) and is divided into the paths $|L\rangle$ and $|R\rangle$, thus achieving the preparation of the pre-selection state, $|\psi_i\rangle$. Next, a weak measurement is implemented to simulate the ITE, which leads to wave packets of $|\psi_i\rangle$ with a minimal shift. The post-selection is composed of $U$, BS2, $X$ and Detectors D1–D3, which achieves the projection operation to the post-selection state, $|\psi_f\rangle$.

In the weak measurement step, the weak values in the quantum Cheshire cat system are extracted based on the theoretical framework of ITE[25]. Under this framework, observable $\hat{A}$ will generate a non-unitary evolution, $U(H,t) = e^{-Ht}$, where $H = \hat{A}$ represents the Hamiltonian of the system, and the interaction time $t$ represents the coupling strength. Here, we define $N_0 = |\langle\psi_i|\psi_f\rangle|^2$ and $N(U) = |\langle\psi_f|U|\psi_i\rangle|^2$, and the normalized incidence rate is given by $N = N(U)/N_0$, which represents the probability of a successful post-selection after the disturbance. As only the real part exists in this experiment, the weak value of observable $\hat{A}$ is $\langle\hat{A}\rangle_w = -(\partial N/\partial t)/2$ (more details are provided in Supplementary Material).

Regarding the post-selection, the post-selection state is set as $|\psi_f\rangle = (|L\rangle|Wave\rangle + |R\rangle|Particle\rangle)/\sqrt{2}$, which is achieved using the quantum gates comprising the operation $U$, BS2, and operator $X$. The action of the operation $U$ realizes the mutual transformation between the wave state and particle state ($|Wave\rangle \to |Particle\rangle$ and $|Particle\rangle \to |Wave\rangle$). Operator $X$ plays a role in the separation of $|Wave\rangle$ and $|Particle\rangle$, i.e., leading $|Wave\rangle$ to Detector D1 and reflecting $|Particle\rangle$ to Detector D3. To verify that the post-selection state projected by the designed setup is indeed $|\psi_f\rangle$, we carry out the deduction from the opposite direction of the quantum state evolution. It is supposed that a wave state $|Wave\rangle$ starts from Detector D1. Afterward, $X$ is injected into BS2 and thus, is naturally divided into the $|L\rangle$ and $|R\rangle$ paths. Therefore, the quantum state can be written as $(|L\rangle|Wave\rangle + |R\rangle|Wave\rangle)/\sqrt{2}$. Consequently, after the operation $U$ is performed on the $|R\rangle$ path, the wave state $|Wave\rangle$ on the $|R\rangle$ path is transformed into the particle state $|Particle\rangle$,

and the state in $|L\rangle$ remains unchanged. Hence, through the above series of operations, the state $|\psi_f\rangle = (|L\rangle|Wave\rangle + |R\rangle|Particle\rangle)/\sqrt{2}$ is post-selected with Detector D1 clicking.

To confirm that the wave and particle attributes have been separated successfully, we need to choose different specific observables $\hat{A}$ for the measurements. Here, these observables are defined to observe the wave or particle attribute constrained in only one path. For example, to observe the particle attribute, we choose the observables $\Pi_P^R = \Pi^R \otimes \Pi_P = |R\rangle\langle R| \otimes |P\rangle\langle P|$ and $\Pi_P^L = \Pi^L \otimes \Pi_P = |L\rangle\langle L| \otimes |P\rangle\langle P|$. To observe the wave property, we choose the observables $\Pi_W^R = \Pi^R \otimes \Pi_W = |R\rangle\langle R| \otimes |W\rangle\langle W|$ and $\Pi_W^L = \Pi^L \otimes \Pi_W = |L\rangle\langle L| \otimes |W\rangle\langle W|$. By applying these operators with Eq. (1), we can obtain the corresponding weak values:

$$\langle \Pi_P^L \rangle_w = 0, \langle \Pi_P^R \rangle_w = \frac{\cos\alpha}{\cos\alpha + \sin\alpha}$$
$$\langle \Pi_W^R \rangle_w = 0, \langle \Pi_W^L \rangle_w = \frac{\sin\alpha}{\cos\alpha + \sin\alpha} \qquad (2)$$

The nonzero weak value of the observable suggests that the system is indeed in a state between the pre- and post-selection states. In contrast, when the weak value equals zero, it reveals that the system is not in that state. From Equation (2), it is evident that the particle attribute is constrained in the $|R\rangle$ path, where as the wave attribute is constrained in the $|L\rangle$ path of the interferometer, which indeed demonstrate the separation of the wave and particle properties. The proportion between the wave and particle attributes in different paths is determined by $\alpha$. For example, when $\alpha = \pi/4$, $\langle \Pi_P^R \rangle_w = \langle \Pi_W^L \rangle_w = 1/2$, which indicates that half of the particle attribute is in the $|R\rangle$ path and half of the wave attribute is in the $|L\rangle$ path.

*Experimental setup.* The experimental setup is shown in Fig. 2. Initially, the single-photon source is generated through the spontaneous parametric down-conversion process[40]. Ultraviolet pulses (approximately 100 mW) with a centre wavelength of 400 nm are employed to pump the type-II β-barium borate (BBO) crystals to attain the photon pairs. One of them is treated as the trigger photon, whereas the other is projected into the core interferometer setup. Due to the optical path difference between the trigger path and the core path, the delay time between the two paths is set as 15 ns, and the coincidence detection window is 3 ns.

First, we prepare the superposition state of the wave and particle attributes with the wave–particle toolbox before separating the wave and particle attributes of the single photon. In this toolbox, the photon passes through a polarization beam splitter (PBS) and a half-wave plate (HWP) whose optical axis is set to α/2. This leads to the preparation of the initial state, $|\psi_0\rangle = \cos\alpha|H\rangle + \sin\alpha|V\rangle$, where $|H\rangle$ and $|V\rangle$ represent the horizontal and vertical polarizations of the photon, respectively. Subsequently, the initial state, $|\psi_0\rangle$, enters a beam displacer (BD), which is a birefringent calcite crystal separating the polarizations of the photons in parallel, i.e., dividing them into the up path with $|H\rangle$ polarization and the down path with $|V\rangle$ polarization. We use the polarizations in different paths to encode the states of the wave and particle attributes. Here, $|H_u\rangle$ and $|V_u\rangle$ denote the horizontal and vertical polarizations in the up path ($|H_d\rangle$ and $|V_d\rangle$ for the down path), respectively. The states of $|Wave\rangle = e^{i\frac{\phi_1}{2}}(\cos\frac{\phi_1}{2}|H_d\rangle - i\sin\frac{\phi_1}{2}|V_d\rangle)$ and $|Particle\rangle = (|H_u\rangle + e^{i\phi_2}|V_u\rangle)/\sqrt{2}$ could be prepared with a set of plates including an HWP and a quarter-wave plate (QWP). These two states are called the "wave" and "particle" states based on whether they exhibit interference with respect to the phase parameters $\phi_1$ or $\phi_2$. More details can be found in Materials and Methods. Considering the expression of the predicted weak value shown in Equation (2), the weak value is only related to the proportion between the wave and particle states (namely, $\alpha$) and independent of $\phi_1$ and $\phi_2$. For simplicity, in our setup, we set $\phi_1 = \phi_2 = 0$. Therefore, the wave state, $|Wave\rangle = |H_d\rangle$, is prepared with an HWP set at 45° and a QWP set at 0°. The particle state, $|Particle\rangle = (|H_u\rangle + |V_u\rangle)/\sqrt{2}$, is prepared with an HWP at 22.5° and a QWP at 45°. Thus, we successfully prepared the superposition input state, $|\psi\rangle = \cos\alpha |Particle\rangle + \sin\alpha |Wave\rangle$ using the wave–particle toolbox.

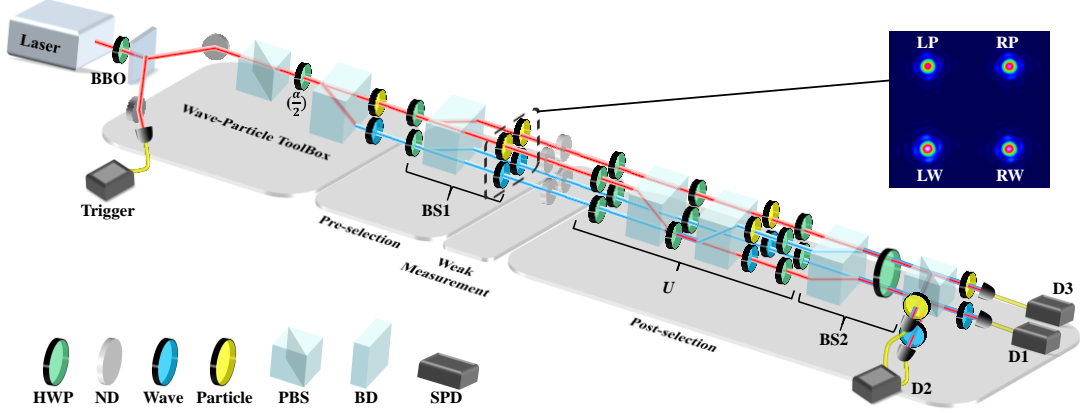

**Fig. 2. Experimental setup for the separation of the wave and particle attributes.** Wave–particle toolbox: A single photon generated through spontaneous parametric down-conversion enters the wave–particle toolbox. The first half-wave plate (HWP) after the polarization beam splitter (PBS) controls the proportion of the horizontal and vertical polarization states ($|H\rangle$ and $|V\rangle$, respectively), namely the initial state, $|\psi_0\rangle = \cos\alpha |H\rangle + \sin\alpha |V\rangle$. A beam displacer (BD) is exploited to divide the beam into the up path (red) encoding the particle attribute and the down path (blue) encoding the wave attribute. The blue and yellow plates comprising a set of plates (HWP and quarter-wave plate) are used to prepare the wave attribute state and particle attribute state, respectively. Pre-selection: The input state is injected into BS1, and the pre-selection state is obtained. Weak measurement: Neutral density (ND) filters with different transmission rates are inserted into the corresponding path to simulate the weak disturbance. Post-selection: The corresponding evolution operation $U$ is only carried out on the right paths. Single-photon detectors (SPD; D1–D3) are used to detect the photons.

Next, the obtained wave–particle superposition state is injected into the improved MZI for further evolution. The three operational steps to achieve the quantum Cheshire cat, which are pre-selection, weak measurement and post-selection, are implemented in sequence in the improved MZI. In the pre-selection stage, the superposition state, $|\psi\rangle$, enters a BS in the left and right paths. The experimental image of the optical modes is also shown in Fig. 2, where R/L represents the right/left side and W/P represents the wave/particle state. Therefore, the pre-selection state, $|\psi_i\rangle = (|L\rangle + |R\rangle)(\cos\alpha |Particle\rangle + \sin\alpha |Wave\rangle)/\sqrt{2}$, is obtained using a BD and a set of wave plates, which realize the role of BS1 in Fig. 1**b**.

Subsequently, in the weak measurement step, a series of neutral density (ND) filters are inserted into different paths of the interferometer to simulate the disturbance in the ITE. The transmission of the ND filter is defined as $T = e^{-2t}$, which relates to interaction time $t$. Therefore, by adjusting the transmission rates, the weak value can be extracted by calculating the slope of the model

In the post-selection step, we initially exchange the particle-like state with the wave-like state through $U$, which is realized using a set of plates and two BDs to exchange the paths and the corresponding polarization states (different colours of light are used in Fig. 2 to illustrate this process more clearly). Finally, a BD, an HWP (optical axis at 22.5°), and a PBS achieve the function of BS2. The function of $X$ in Fig. 1**b**, i.e., ensuring that only the wave state $|Wave\rangle$ passes to Detector D1 and that particle state $|Particle\rangle$ reflects to Detector D3, is achieved naturally in our encoding. SPDs D1–D3 are used to detect the photons, whose signals coincide with the trigger.

*Experimental results.* First, to verify the high performance of our experimental setup and ensure that the obtained results are convincing, quantum state tomography is performed for the output state of BS2. The average fidelity of all the reconstructed density matrices of the output state is 99.45 ± 0.26%, which manifests the high interference visibility of the employed setup. The quantum states tomography results for all degrees can be found in Supplementary Material.

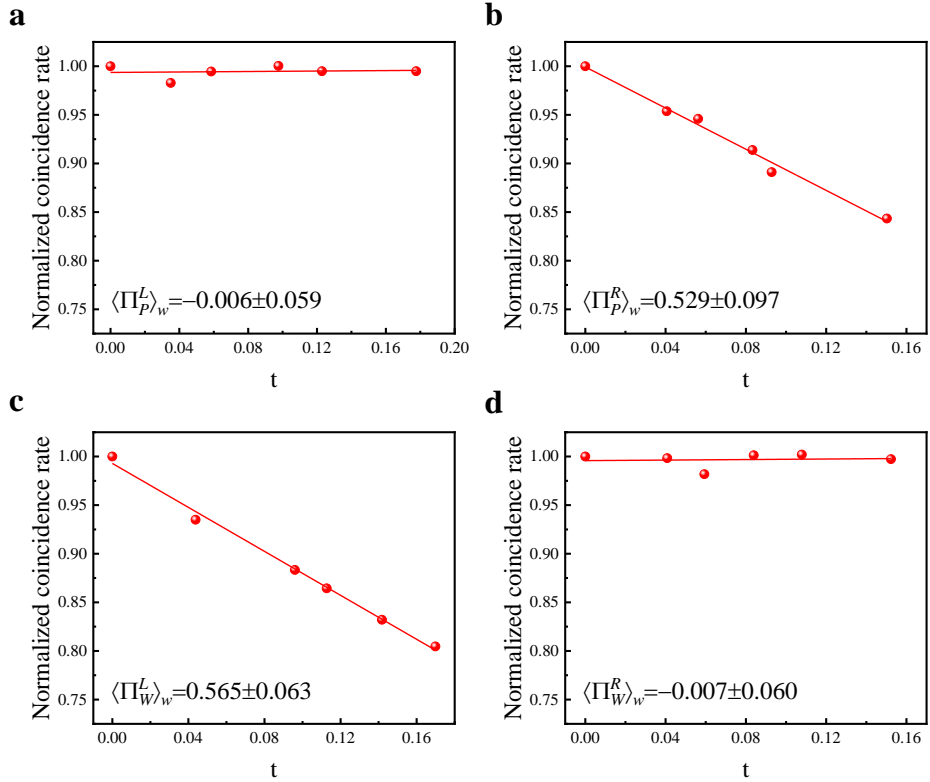

**Fig. 3. Experimental data when inserting a series of ND filters with different transmission rates for different observables with α ≈ 45°. a-d** Points with error bars (overly small to be visible) represent the experimental results and are fitted with lines for $\langle\Pi_P^L\rangle_w$, $\langle\Pi_P^R\rangle_w$, $\langle\Pi_W^L\rangle_w$, and $\langle\Pi_W^R\rangle_w$. The corresponding weak value is displayed in the lower-left corner of each figure.

When implementing the extraction of the weak values based on ITE, as introduced above, a series of ND filters with different transmission rates is employed to simulate the perturbation. A detailed illustration can be found in Materials and Methods. After repeating this procedure with many transmission rates, a curve can be obtained by linear fitting based on these data points. According to the derivation mentioned before, a minus half of the slope is the expected weak value.

The experimental data are shown in Fig. 3 for the case where α is ~45° (results for other degrees are shown in the Supplementary Material), and the corresponding weak value obtained after processing is marked in each figure. The weak values are theoretically zero for observables $\Pi_P^L$ and $\Pi_W^R$, and the corresponding experimental results shown in Figs. 3**a, d** match the theoretical predictions very well, indicating that there are no particle (wave) properties constrained in the left (right) path. For observable $\Pi_P^R$ with the experimental results shown in Fig. 3**b**, half of the particle attribute is in the right path. The observable $\Pi_W^L$ shown in Fig. 3**c** demonstrates that half of the wave property is in the left path. The experimental error bars are the standard deviations estimated according to Monte Carlo methods with photon counting events following the Poisson distribution.

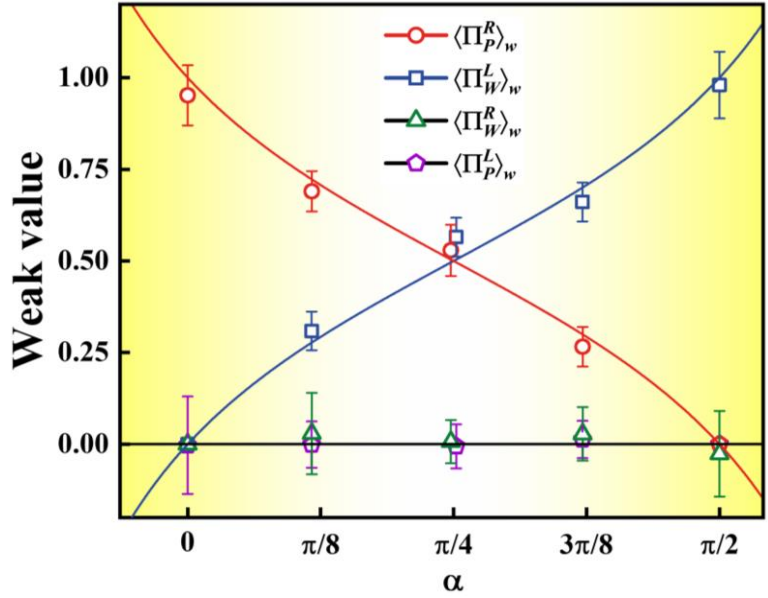

**Fig. 4 Comparison between the theoretical prediction and experimental results.** The curves with different colours represent the corresponding theoretical weak value, and points of different shapes represent the groups of experimental data. Error bars for α are extremely small and near invisible.

In our experiment, we set different angles for α from 0 to π/2 to show the function curve of different weak values for the observables, and the experimental results are shown in Fig. 4. For every data point, the acquisition method is the same as the above-described procedure for α ≈ 45°. As mentioned earlier, a nonzero weak value of the observable suggests that the system is indeed in the state between the pre- and post-selection states. As shown in Fig. 4, the weak values of the observables $\Pi_P^R$ and $\Pi_W^L$ prove undoubtedly that the wave and particle attributes have indeed been separated, and the proportion of the wave/particle-like states is dependent on α. The obtained experimental data are highly consistent with the theoretical prediction, which confirms that we successfully separated the wave and particle attributes of a single photon under the quantum Cheshire cat framework experimentally.

**Discussion**

In this proof-of-principle work, we experimentally demonstrate the theoretical construction of separating the wave and particle attributes of a physical entity for the first time. Under the framework inspired by the quantum Cheshire cat and with the help of ITE, weak values that confirm the successful separation are obtained by selecting appropriate pre-selection and post-selection states.

Finally, our work improves the comprehesion of wave–particle duality and will inspire numerous theoretical and experimental works in this field. It also provides a reference for the experiment on tripartite separation: the wave attribute, particle attribute and the physical entity itself, i.e., the so-called quantum Cheshire "supercat[39]."

For potential application prospects, our work may provide references for quantum precision measurements[41] and enhance the understanding of counterfactual communication[42]. Precision measurement requires performing a measurement only for one mechanical quantity and does not disturb other variables. This could offer a broad perspective for multiparameter estimation[43, 44] through the separation of these observables, which is a vital problem in quantum metrology. Counterfactual communication proposed in recent years, which describes a phenomenon in which communication can be achieved even without physical particle transmission, has attracted widespread interest and has been proven experimentally[45]. However, its core principle remains unclear[46], and control of the property without the particle itself has not been experimentally achieved thus far[47]. Our work provides an experimental platform for achieving this goal.

**Materials and Methods**
**Details of the encoding method for the wave–particle state**

The description of the wave/particle attribute was first proposed in Wheeler's delayed-choice *Gedanken experiment*[7], which has been widely adopted in several influential works[9-11].

Here, the same encoding method for the waves/particle attribute is implemented, as shown in Fig. 1**a**. For example, $|Wave\rangle = e^{i\frac{\phi_1}{2}}(cos\frac{\phi_1}{2}|0\rangle - i\,sin\frac{\phi_1}{2}|1\rangle)$ and $|Particle\rangle = (|0\rangle + e^{i\phi_2}|1\rangle)/\sqrt{2}$, where $|0\rangle/|1\rangle$ represents the two paths in the MZI. If BS2 exists, by changing the phase difference $\phi_1$ between two paths, the probability of receiving a single photon in the detector will change with $\phi_1$ due to the interference. This indicates that the photon has travelled both arms of the MZI, leading to the interference phenomenon with the wave attribute. If BS2 is removed, the probability that the detector on each output receives a single photon will always be 1/2 regardless of the phase difference $\phi_2$ between the two paths. This indicates that the photon must have passed one of the two paths in the interferometer, revealing its particle nature. In our work, the polarization degree of freedom is employed to replace the path degree of freedom. Path states $|0\rangle/|1\rangle$ are actually converted into polarization states $|H\rangle/|V\rangle$ in up or down paths. Thus, these two states are written as $|Wave\rangle = e^{i\frac{\phi_1}{2}}(cos\frac{\phi_1}{2}|H_d\rangle - i\,sin\frac{\phi_1}{2}|V_d\rangle)$ and $|Particle\rangle = (|H_u\rangle + e^{i\phi_2}|V_u\rangle)/\sqrt{2}$. The utilization of the polarization states for encoding simplifies the experimental setup, where only one path is required.

**Details of the method used to extract the weak value**

For the detailed process of extracting the weak value, we need to implement the following operations. First, the total count of the photons in Detector D1 is recorded as $N_0$ in the formula $N = N(U)/N_0$. Thereafter, we insert an ND filter in the related arm of the interferometer to extract the weak value of the corresponding observer. Now, the count of the photons in D1 will change because of this operation and is recorded as $N(U)$. Thus, the ordinate $N = N(U)/N_0$ of the data point can be obtained. The abscissa t is calculated through $T = e^{-2t}$, where T is the transmission rate of the ND filter, which can be measured easily in this experimental setup. The obtained data point is the same as that shown in Fig 3 in the main text. After repeating the above procedure using ND filters with different transmission rates (here, we repeat this procedure five times), a series of data points after normalization is obtained. Finally, by the linear fitting of these data points using the least-squares method, a line with a specific slope is obtained. Based on Equation (3) in the Supplementary Material, a minus half of the slope is the expected weak value.

**Acknowledgements.**
This work was supported by the Innovation Program for Quantum Science and Technology (Nos. 2021ZD0301200, 2021ZD0301400), National Natural Science Foundation of China (Grant Nos. 11821404, 61725504, U19A2075, 61975195, 11875167, 12075001), Anhui Initiative in Quantum Information Technologies (Grant No. AHY060300), and Fundamental Research Funds for the Central Universities (Grant No. WK2030380017).

**Conflict of interest.**
The authors declare no competing interests.

**Author contributions.**
J.-K.L. designed and performed the experiment with the help of K.S., Y.W., Z.-Y.H. J.-K.L. analysed the data with the help of K.S. and Z.-H.L. J.-K.L. wrote the paper. J.Z., X.-Y.F. and J.-L.C. supervised the theoretical part of the project. J.-S.X., C.-F.L. and G.-C.G. supervised the project. All authors read the paper and discussed the results.